\documentclass[preprint,showpacs, preprintnumbers,amsmath,amssymb,nofootinbib]{revtex4}
\usepackage{pifont}
\usepackage{amssymb}

\usepackage{graphicx}
\usepackage{dcolumn}
\usepackage{bm}

\begin{document}
\title{Constraints on the generalized tachyon field models from latest observational data}

\author{Rong-Jia Yang$^{1}$\footnote{yangrj05@mails.tsinghua.edu.cn}, Shuang Nan Zhang$^{1,2,3}$, and Yuan
Liu$^{1}$}
\address{$^{1}$ Department of Physics and Tsinghua Center for Astrophysics,
Tsinghua University, Beijing 100084, China
\\ $^{2}$ Key Laboratory
of Particle Astrophysics, Institute of High Energy Physics, Chinese Academy of
Sciences, P.O. Box 918-3, Beijing 100049, China,
\\$^{3}$Physics Department, University of Alabama in Huntsville,
Huntsville, AL 35899, USA}


\begin{abstract}
We consider constraints on generalized tachyon field (GTF) models
from latest observational data (including 182 gold SNIa data, the
shift parameter, and the acoustic scale). We obtain at $68.3\%$
confidence level $\Omega_{\rm m}=0.37\pm0.01$,
$k_0=0.09^{+0.04}_{-0.03}$, $\alpha=1.8^{+7.4}_{-0.7}$ (the best-fit
values of the parameters) and $z_{q=0}\sim 0.47-0.51$ (the
transitional redshift) for GTF as dark energy component only;
$k_0=0.21^{+0.20}_{-0.18}$, $\alpha=0.57\pm0.01$ and $z_{q=0}\sim
0.49-0.68$ for GTF as unification of dark energy and dark matter. In
both cases, GTF evolves like dark matter in the early universe. By
applying model-comparison statistics and test with independent
$H(z)$ data, we find GTF dark energy scenario is favored over the
$\Lambda$CDM model, and the $\Lambda$CDM model is favored over GTF
unified dark matter by the combined data. For GTF as dark energy
component, the fluctuations of matter density is consistent with the
growth of linear density perturbations. For GTF unified dark matter,
the growth of GTF density fluctuations grow more slowly for
$a\rightarrow1$, meaning GTF do not behave as classical $\Lambda$CDM
scenarios.

\end{abstract}



\maketitle

\section{Introduction}
Tachyon field can be seen as special cases of k-essence \cite{sen}
and has been explored extensively \cite{Padmanabhan, Gorini, Ying,
Choudhury, Hao, Bagla, Garousi, Garousi04, Copeland, Cardenas, Shiu,
Wang06, Koshelev}. For a constant potential, the tachyon field can
be generalized as
\begin{eqnarray}
\label{1}F(X)=-V_0(1-2X^{n})^{\frac{1}{2n}},
\end{eqnarray}
called generalized tachyon field (GTF) \cite{Chim04}, where $n$ is a non-zero
parameter. Such model can be considered as a scalar field realization of the
generalized Chaplygin gas (GCG) \cite{Frolov, Kame01, Chim04, Gorini05}. With the
theoretical constraint on purely kinetic k-essence: $F_{x}=F_{0}a^{-3}$, where $F_{0}$
is a constant \cite{Chim04, Sch04, Yang}, one gets the expressions for the equation of
state parameter (EoS) $w_{\rm k}$ and the sound speed $c^{2}_{\rm s}$ of the GTF
depending on the scale factor (so the redshift) respectively
\begin{eqnarray}
\label{18}w_{\rm k}&=&-\frac{1}{1+2k_0^{2\alpha}(1+z)^{6\alpha}},\\
\label{19}c^{2}_{\rm s}&=&-(2\alpha-1)w_{\rm k},
\end{eqnarray}
where $\alpha=n/(2n-1)$ and $k_0$ is a constant
($-\infty<k_0<+\infty$, but because of the exponent 2, the case
$k_0\geqslant 0$ and the case $k_0\leqslant 0$ are equivalent).
Obviously, the EoS parameter is negative and not less than $-1$,
meaning that the GTF does not violate the weak energy condition. For
$k_0=0$, the EoS reduces to $-1$; that is to say, the $\Lambda$CDM
model is contained in the GTF dark energy scenario as one special
case. As Eq. (\ref{19}) shows, $\alpha<1/2$ will lead to imaginary
sound speed and thus instabilities \cite{Gar99}, so we will only
concentrate on the case of $\alpha>1/2$ in the following. In this
case, the behavior of the EoS (\ref{18}), being $\simeq-0$ in the
early Universe, runs closely to $-1$ in the future for $k_0\neq 0$.
Such behavior can, to a certain degree, solve the fine-tuning
problem \cite{Arm00, Chi}.

There have been a number of papers considering observational
constraints on GCG model, such as Refs. \cite{Bean2003, Gorini03,
Colistete2003, Sen05, Multamaki2004, Makler2003a, Silva2003,
Cunha2004, Bertolami2004, Zhu2004, Zhang, Bouh05, Gong,
Giannantonio, Biesiada, Jimeez, Ghosh2007, Lobo2006, wu, Bento2003a,
Alcaniz, Bertolami2006, Dev2003, Chen2003a, Bilic2002, Alcaniz2003,
Sandvik, Perrotta, Amendola05}. As the scalar field realization of
GCG, GTF with Lagrangian (\ref{1}) yet has not been fully analyzed
with observational data currently available. This is necessary if
such exotic types of matter are to be considered as serious
alternatives to the $\Lambda$CDM scenario. Cosmological models that
include (generalized) Chaplygin gas component can be divided into
two classes: models with and without a significant CDM component. It
now appears increasingly likely from both theoretical stability
issues and observational constraints (e.g. \cite{Bean2003, Sandvik,
Perrotta, Amendola05}) from matter clustering properties (dark
matter is very clumpy while dark energy is quite smooth out to the
Hubble scale) that dark matter and dark energy are not the same
substance. Also it appears rather difficult to unify dark matter and
dark energy into a single scalar field in the context of the string
landscape \cite{Liddle}.

Nevertheless, in this paper we will consider these two cases: GTF as
dark energy only and as unification of dark matter and dark energy,
without loss of generality. The data sets used here include the
recently released 182 gold supernova (SNIa) data \cite{rie06}, the
shift parameter $R$ and the acoustic scale $l_{\rm a}$ from
observations of CMB \cite{wang}. Our results show that GTF dark
energy scenario is favored over the $\Lambda$CDM model, and the
$\Lambda$CDM model is favored over GTF as unification of dark matter
and dark energy by the combined data.

\section{The luminosity distance of the GTF model}
For a flat and homogeneous Friedmann-Robertson-Walker (FRW) space,
the Einstein's field equations take the forms:
\begin{eqnarray}
\label{5}H^{2}:=\left (\frac{\dot{a}}{a}\right)^2=H^2_0E^2.
\end{eqnarray}
For GTF as dark energy component only
\begin{eqnarray}
\label{2}E(\Omega_{\rm m}, k_0, \alpha)=[\Omega_{\rm m} (1+z)^3+\Omega_{\rm
r}(1+z)^4+(1-\Omega_{\rm m}-\Omega_{\rm r})f(z)]^{1/2},
\end{eqnarray}
where $\Omega_{\rm m}$ and $\Omega_{\rm r}$ are the present
dimensionless density parameters of matter (including both the dark
and baryonic matter) and radiation respectively; $f(z)$ is the ratio
of the energy density of GTF with respect to its present value
$f(z)\equiv \rho_{\rm k}(z)/\rho_{\rm
k}(0)=\exp[3\int^{1}_a\frac{da'}{a'}(1+w_{\rm k}(a'))]$. For GTF as
unification of dark matter and dark energy
\begin{eqnarray}
\label{3}E(k_0, \alpha)=[\Omega_{\rm b} (1+z)^3+\Omega_{\rm
r}(1+z)^4+ (1-\Omega_{\rm b}-\Omega_{\rm r})f(z)]^{1/2},
\end{eqnarray}
where $\Omega_{\rm b}$ is the present dimensionless density parameter of baryonic
matter. The Hubble-parameter free luminosity distance is expressed as
\begin{equation}
\label{8}D_{\rm L}(z)=H_0(1+z)\int^{z'}_0 \frac{dz'}{H}.
\end{equation}

\section{Observational constraints and the evolution of the GTF}
To consider the  best fit values of the parameters, we study
observational bounds on the GTF models for a flat universe. Our
constraints come from combinations of $182$ gold supernova data
\cite{rie06} and the CMB observation \cite{wang}.

The SNIa data which provide the main evidence for the existence of
dark energy in the framework of standard cosmology \cite{rie98}.
Here we use a recently published dataset consisting of $182$ SNIa
with 23 SNIa at $z\gtrsim 1$ obtained by imposing constraints
$A_{\rm v}<0.5$ (excluding high extinction) \cite{rie06}. Each
data point at redshift $z_i$ includes the Hubble-parameter free
distance modulus $\mu_{\rm obs}(z_i)$ ($\equiv m_{\rm obs}-M$,
where $M$ is the absolute magnitude) and the corresponding error
$\sigma^2(z_i)$. The resulting theoretical distance modulus
$\mu_{\rm th}(z)$ is defined as
\begin{eqnarray}
\label{20}\mu_{\rm th}(z)\equiv 5\log_{10}D_{\rm L}(z)+\mu_0,
\end{eqnarray}
where $\mu_0\equiv 5\log_{10}h-42.38$ is the nuisance parameter
which can be marginalized over \cite{per05}. Fitting $\Lambda$CDM
model with these $182$ SNIa data, the best-fit value of parameter is
$\Omega_{\rm m}=0.34$; fitting GCG as dark energy component, it is
$\Omega_{\rm m}=0.39$ \cite{Sen05}.

In order to break the degeneracies among the parameters, we consider
the shift parameter $R$ and the acoustic scale $l_{\rm a}$
\cite{Pag03} which are nearly uncorrelated with each other and
defined as
\begin{eqnarray}
R&\equiv&\Omega^{1/2}_{\rm m}\int^{z_{\rm CMB}}_0\frac{dz}{E(z)},\\
l_{\rm a}&\equiv&\frac{\pi \int^{z_{\rm CMB}}_0
dz/E(z)}{\int^{a_{\rm CMB}}_0c_{\rm s}da/(a\dot{a})}.
\end{eqnarray}
For the case of GTF as dark energy only, $\Omega_{\rm r}/\Omega_{\rm
m}=1/(1+z_{\rm eq}) (z_{\rm eq}=2.5\times 10^4 \Omega_{\rm m} h^2
(T_{\rm CMB}/2.7 {\rm K})^{-4})$ with the redshift of recombination
$z_{\rm CMB}=1089$ $(a_{\rm CMB}=1/[1+z_{\rm CMB}])$. The sound
speed is $c_{\rm s}=1/\sqrt{3(1+R_{\rm b}a)}$ with $R_{\rm
b}a=31500\Omega_{\rm b}h^2(T_{\rm CMB}/2.7 {\rm K})^{-4}a$. COBE
four year data give $T_{\rm CMB}=2.728$ K \cite{Fix96}. For the case
of GTF as unification of dark matter and dark energy, $\Omega_{\rm
m}=\Omega_{\rm b}+(1-\Omega_{\rm b}-\Omega_{\rm r})(1+w_{\rm
k0})^{1/2\alpha}$ with $w_{\rm k0}=-1/(1+2k^{2\alpha}_{0})$ is the
effective matter density parameter \cite{wu, Amendola05}, and
$\Omega_{\rm r}=10^{-5}$ is assumed. The three-year WMAP data give
$\Omega_{\rm b}h^2=0.022\pm0.00082$, $R=1.70\pm0.03$ and $ l_{\rm
a}=302.2\pm1.2$ \cite{wang}. Here we use the acoustic scale $l_{\rm
a}$ with a prior of $H_0=62.3\pm 1.3$ (random )$\pm5.0$
(systematic)(km/s) Mpc$^{-1}$ from HST Cepheid-calibrated luminosity
of Type Ia SNIa observations recently \cite{san06}.

The shift parameter $R$ is a geometrical measure as it measures the size of apparent
sound horizon at the epoch of recombination. Keeping the sound horizon size fixed,
different cosmological models lead to different background expansion and hence the
shift parameter can be used to compare and constrain different models. However, the
sound horizon size also changes when varying cosmological parameters, most notably
changing the matter density $\Omega_{\rm m}$. Hence in general the shift parameter will
not be an accurate substitute for CMB dada, but the combination of the shift parameter
$R$ and the acoustic scale $l_{\rm a}$ has been proved to be a good and efficient
approximation to the full CMB data to probe cosmological models \cite{wang, elg, wri}.

Since the SNIa, the shift parameter $R$, and the acoustic scale $l_{\rm a}$ are
effectively independent measurements, we can simply minimize their total $\chi^{2}$
value given by \cite{Mov06, Wu06, Wei07}
\begin{eqnarray}
\label{21}\chi^2(\Omega_{\rm m}, k_0, \alpha)=\chi^{2}_{\rm
SNIa}+\chi^{2}_{\rm R}+\chi^{2}_{\rm l_a},
\end{eqnarray}
where
\begin{eqnarray}
\chi^{2}_{\rm SNIa}&=&\sum^{N}_{i=1}\frac{(\mu^{\rm obs}_{\rm
L}(z_i)-\mu^{\rm th}_{\rm L}(z_i))^2} {\sigma^2_i},\\
\chi^{2}_{\rm R}&=&\left(\frac{R-1.70}{0.03}\right)^2,
\end{eqnarray}
and
\begin{eqnarray}
\chi^{2}_{\rm l_a}=\left(\frac{l_{\rm a}-302.2}{1.2}\right)^2,
\end{eqnarray}
in order to find the best fit values of the parameters of the GTF
models.

\subsection{The case of GTF as dark energy only}
For the case of GTF as dark energy component only, we obtain the best fit values of the
parameters at $68\%$ confidence level: $\Omega_{\rm m}=0.37\pm0.01$,
$k_0=0.09^{+0.04}_{-0.03}$ and $\alpha=1.8^{+7.4}_{-0.7}$ with $\chi^2_{\rm k,
min}=159.30$ ($p(\chi^2>\chi^2_{\rm k, min})=0.88$), comparing with $\Omega_{\rm
m}=0.39\pm0.009$ and $\chi^2_{\rm \Lambda, min}=168.59$ ($p(\chi^2>\chi^2_{\rm \Lambda,
min})=0.77$) in the $\Lambda$CDM case. The probability of the improvement in the
$\chi^2_{\rm min}$ by chance is $0.59\%$ with F-statistic value of $5.28$ resulted from
F-test.

Now we apply information criteria to assess the strength of models.
These statistics favor models that give a good fit with data. In
this paper we use the Akaike Information Criterion (AIC)
\cite{Akaike} and the Bayesian Information Criterion (BIC)
\cite{Schwarz} (see also \cite{Liddle04} and reference therein) to
select the best-fit models. Comparing with the $\Lambda$CDM case,
the difference of the Akaike Information Criterion (AIC) is $\Delta
$AIC$=-5.29$, supporting GTF dark energy scenario; the Bayesian
Information Criterion (BIC) is $\Delta$BIC$=1.14$, less supporting
GTF dark energy scenario.

\begin{figure}
\includegraphics[width=10cm]{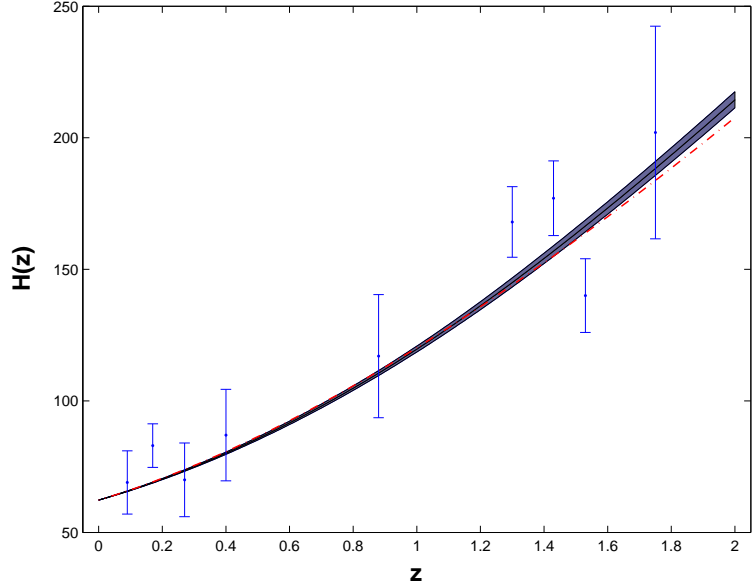}
\caption{The predicted values of the Hubble parameter $H$ of the GTF
as dark energy only in $68.3\%$ confidence level limits from fitting
the combined data, compared with the observational $H(z)$ data with
error bars and the $\Lambda$CDM case (the dash-dot line).
\label{Fig4}}
\end{figure}

Because model-comparison statistics can not discriminate between GTF
dark energy scenario and the $\Lambda$CDM model. We carry out
another independent observational test with 9 $H(z)$ data points
\cite{r8, r9} in the range $0\lesssim z\lesssim 1.8$ obtained by
using the differential ages of passively evolving galaxies
determined from the Gemini Deep Deep Survey (GDDS) \cite{r10} and
archival data \cite{r11, r12}. We compare these observational $H(z)$
data with the predicted values of the Hubble parameter $H$ of the
GTF dark energy scenario for the case of ($\Omega_{\rm m}=0.37$,
$k_0=0.09$, $\alpha=1.8$) and the case of ($\Omega_{\rm m}=0.39$,
$k_0=0$) respectively. We find $\chi^2=11.87$
($p(\chi^2>11.86)=0.22$) for the former case and $\chi^2=12.66$
($p(\chi^2>12.66)=0.18$) for the latter case, both with $9$ degrees
of freedom because no fitting is done with the $H(z)$ data. This
serves as an independent evidence that the GTF dark energy scenario
is favored over the $\Lambda$CDM model by these $H(z)$ data. The
predicted values of the Hubble parameter $H$ of the GTF dark energy
scenario in $68.3\%$ confidence level limits compared with the
observational $H(z)$ data is shown in figure \ref{Fig4}; the
$\Lambda$CDM case is also presented for comparison.

\begin{figure}
\includegraphics[width=10cm]{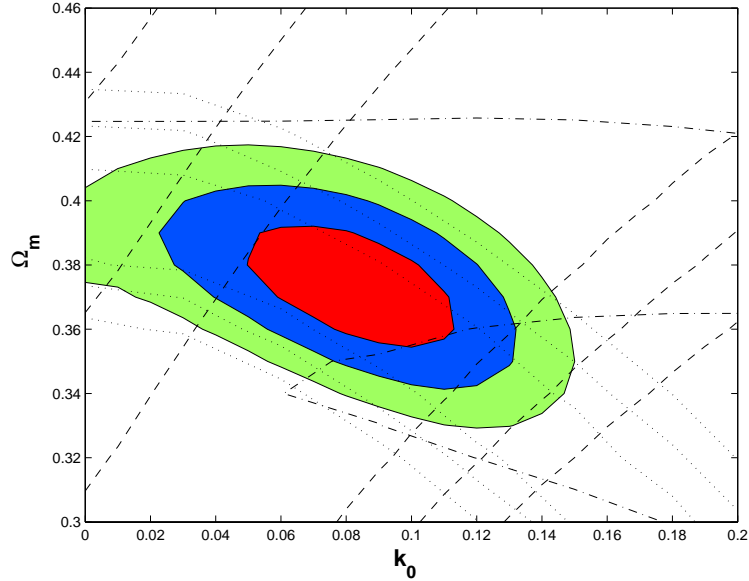}
\caption{The $68.3\%$, $95.4\%$ and $99.7\%$ confidence regions in the
$k_0$-$\Omega_{\rm m}$ plane with $\alpha$ at its best-fit value of $1.8$, for the case
of GTF as dark energy only. The dot-dashed lines, dotted lines, dashed lines represent
the results from the 182 gold SNIa sample, the acoustic scale and the shift parameter
respectively. The colored areas show the results from the combination of these three
data sets. \label{Fig1}}
\end{figure}

\begin{figure}
\includegraphics[width=10cm]{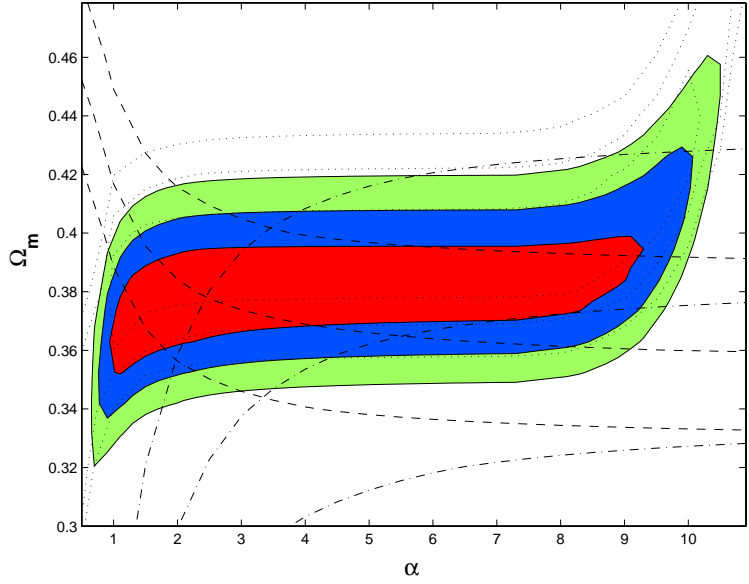}
\caption{The same confidence regions as in Fig \ref{Fig1} in the $\alpha$-$\Omega_{\rm
m}$ plane with $k_0$ at its best-fit value of $0.09$, for the case of GTF as dark
energy only. \label{Fig2}}
\end{figure}

\begin{figure}
\includegraphics[width=10cm]{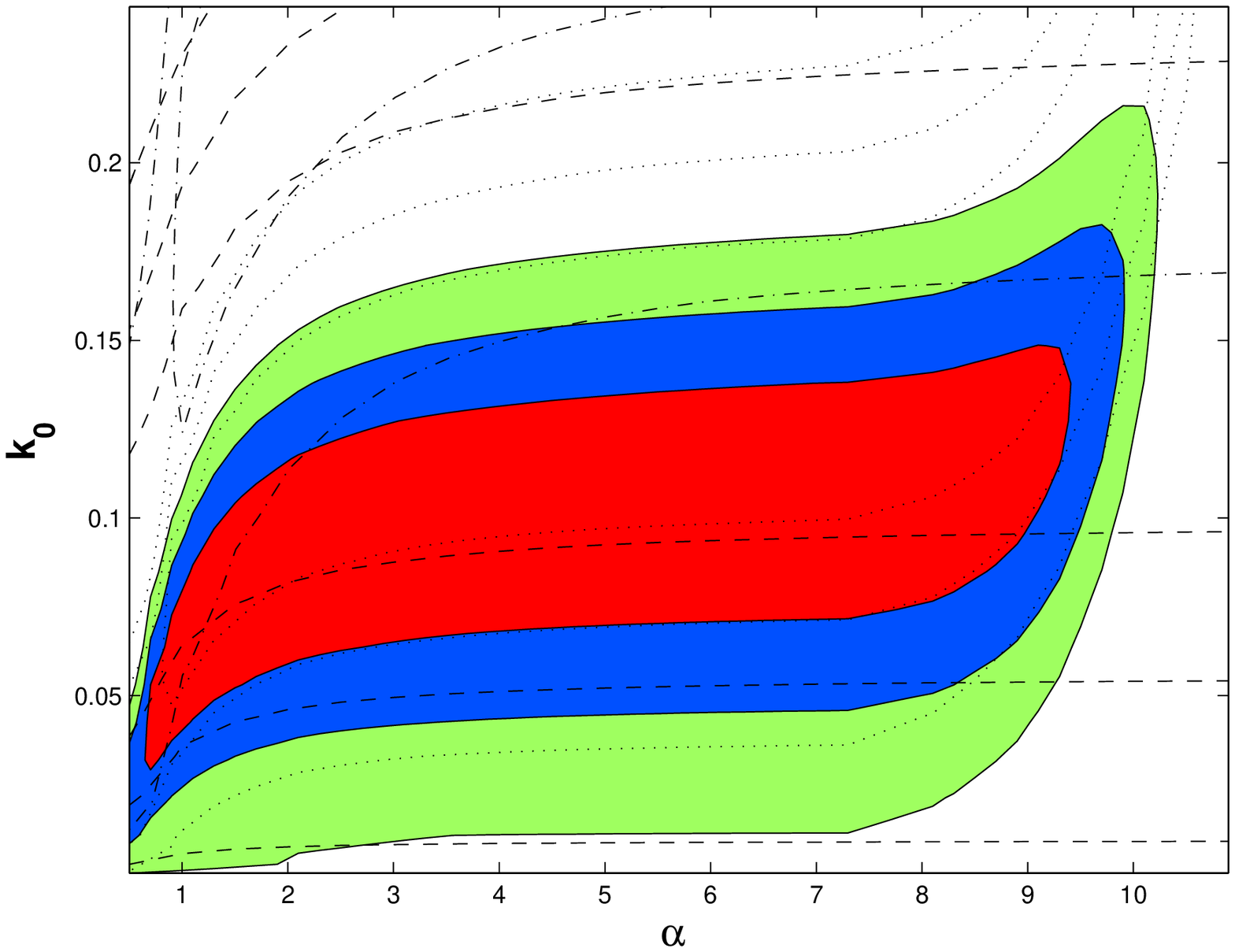}
\caption{The same confidence regions as in Fig \ref{Fig1} in the $\alpha$-$k_0$ plane
with $\Omega_{\rm m}$ at its best-fit value of $0.37$, for the case of GTF as dark
energy only. \label{Fig3}}
\end{figure}

Figures \ref{Fig1}, \ref{Fig2}, \ref{Fig3} show the $68.3\%$,
$95.4\%$ and $99.7\%$ joint confidence contours in the $\Omega_{\rm
m}$-$k_{0}$ plane with $\alpha$ at its best fit value of $1.8$, the
$\Omega_{\rm m}$-$\alpha$ plane with $k_0$ at its best fit value of
$0.09$, and the $\alpha$-$k_{0}$ plane with $\Omega_{\rm m}$ at its
best fit value of $0.37$ respectively. The dot-dashed lines, dotted
lines, dashed lines represent the results from the $182$ gold SNIa
sample, the acoustic scale $l_{\rm a}$ and the shift parameter $R$
respectively. The colored areas show the results from the
combination of these three data sets. Obviously the current
observational bounds on the index $\alpha$ are considerably weak.

\subsection{The case of GTF as unification of dark matter and dark energy}
For the case of GTF as unification of dark matter and dark energy, we find the best fit
values of the parameters at $68\%$ confidence level: $k_0=0.21^{+0.2}_{-0.18}$ and
$\alpha=0.57\pm0.01$ with $\chi^2_{\rm k, min}=167.27$ ($p(\chi^2>\chi^2_{\rm
k,min})=0.78$).

For GTF as unification of dark matter and dark energy, $k_0=0$ dose not correspond to
the $\Lambda$CDM case, so we can not apply F-test \cite{Protassov} for model selection,
but we can still apply AIC and BIC. Comparing with the $\Lambda$CDM case, we find
$\Delta$AIC$=0.68$ and $\Delta$BIC$=3.89$. Comparing with the case of GTF as dark
energy, we find $\Delta$AIC$=5.97$ and $\Delta$BIC$=2.8$. These results of
model-comparison statistics indicate that the case of GTF as unification of dark matter
and dark energy is not favored by the combined data.

\begin{figure}
\includegraphics[width=10cm]{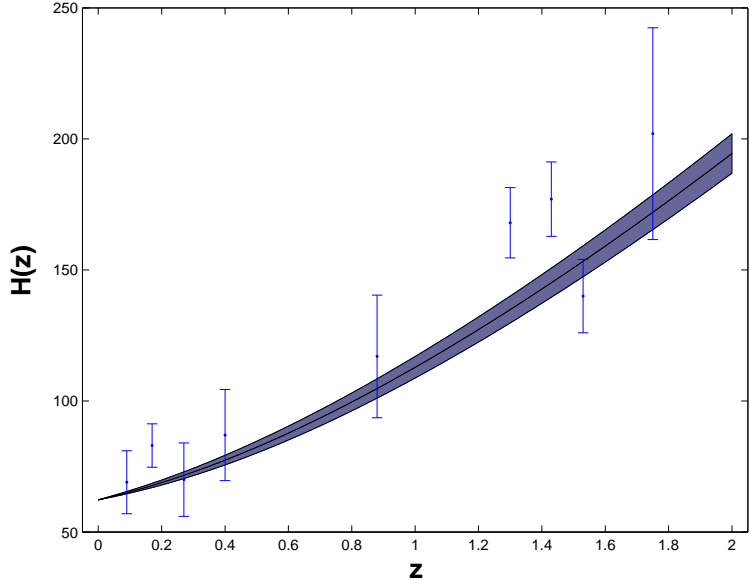}
\caption{The predicted values of the Hubble parameter $H$ of GTF
unification of dark matter and dark energy in $68.3\%$ confidence
level limits from fitting the combined data, compared with the
observational $H(z)$ data with error bars. \label{Fig13}}
\end{figure}

To confirm this result, we also carry out the independent 9 $H(z)$
data points \cite{r8, r9} test. We find $\chi^2=16.60$
($p(\chi^2>11.86)=0.06$), meaning that GTF as unification of dark
matter and dark energy is also not favored by these $H(z)$ data as
shown in figure \ref{Fig13}.

\begin{figure}
\includegraphics[width=10cm]{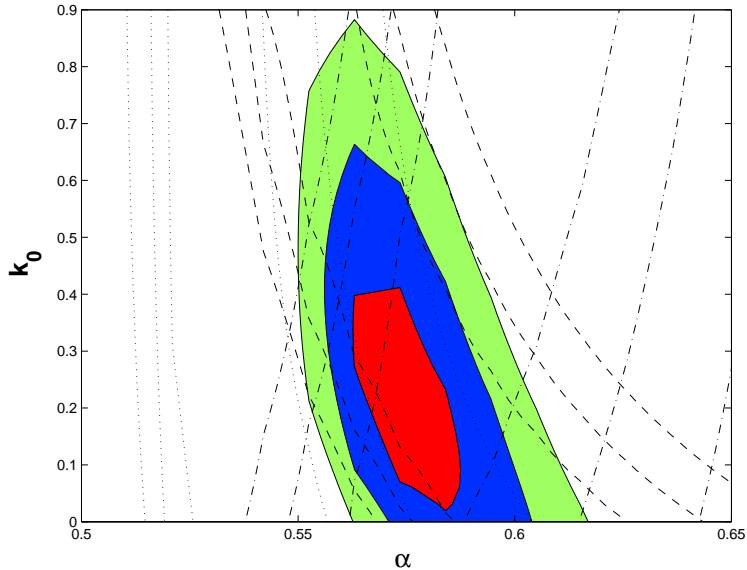}
\caption{The $68.3\%$, $95.4\%$ and $99.7\%$ confidence regions in the $\alpha$-$k_{0}$
plane, for the case of GTF as unification of dark matter and dark energy. The
dot-dashed lines, dotted lines, dashed lines represent the results from the 182 gold
SNIa sample, the shift parameter, and the acoustic scale respectively. The colored
areas show the results from the combination of these three data sets. \label{Fig8}}
\end{figure}

Figure \ref{Fig8} shows the $68.3\%$, $95.4\%$ and $99.7\%$ joint confidence contours
in the $\alpha$-$k_{0}$ plane. The dot-dashed lines, dotted lines, dashed lines
represent the results from the $182$ gold SNIa sample, the shift parameter $R$ and  the
acoustic scale $l_{\rm a}$ respectively. The colored areas show the results from the
combination of these three data sets. Obviously the current observational bounds on the
index $k_{0}$ are considerably weak.

\subsection{The evolution of the GTF}
To study the evolution of the GTF, we investigate the deceleration
parameter $q(z)$, the EoS parameter $w_{\rm k}(z)$, and the energy
density $\rho_{\rm k}(z)$. For GTF as dark energy component alone,
the deceleration parameter $q(z)$ is defined as
\begin{eqnarray}
q(z)=-a\ddot{a}/\dot{a}^2=\frac{1}{2}\Omega_{\rm
m}(z)+\frac{1+3w_{\rm k}(z)}{2}\Omega_{\rm k}(z),
\end{eqnarray}
where $\Omega_{\rm k}$ is energy density parameter of GTF. For GTF
as unification of dark matter and dark energy, the deceleration
parameter $q(z)$ is given by
\begin{eqnarray}
q(z)=\frac{1}{2}\Omega_{\rm b}(z)+\frac{1+3w_{\rm
k}(z)}{2}\Omega_{\rm k}(z),
\end{eqnarray}
Because we only consider the evolution of the deceleration parameter
at low redshift, the radiation is ignored here.

\begin{figure}
\includegraphics[width=10cm]{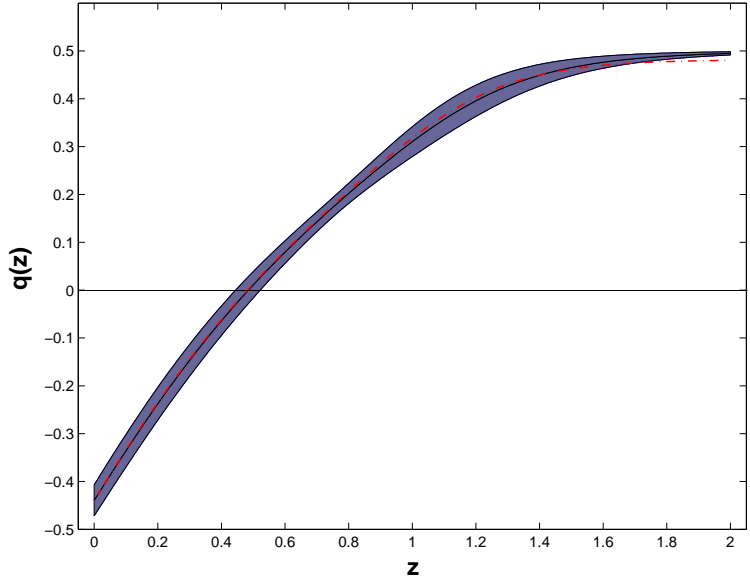}
\caption{The deceleration parameter as a function of redshift in $68.3\%$ confidence
level limits from fitting the combined data, compared with the $\Lambda$CDM case (the
dash-dot line), for GTF as dark energy component only. \label{Fig5}}
\end{figure}

\begin{figure}
\includegraphics[width=10cm]{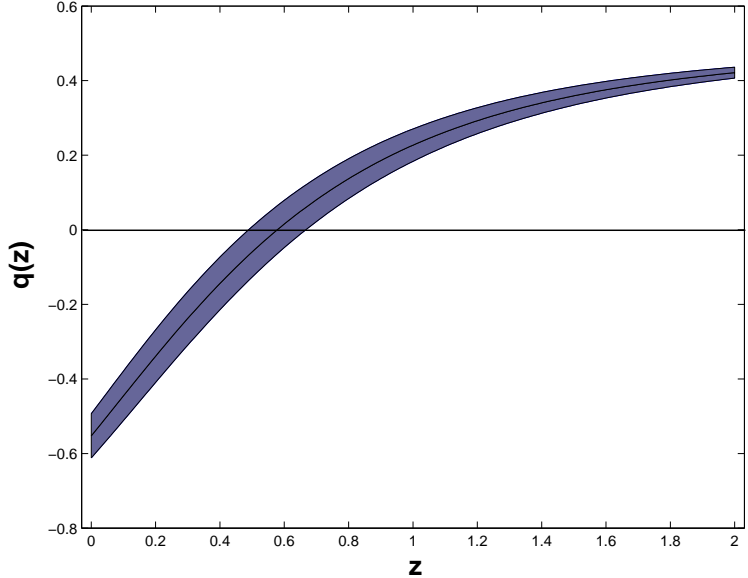}
\caption{The deceleration parameter as a function of redshift in
$68.3\%$ confidence level limits from fitting the combined data, for
GTF as unification of dark matter and dark energy. \label{Fig9}}
\end{figure}

For the case of GTF as dark energy component only, the present value of the
deceleration parameter $q(z)$ is found to be $-q_{z=0}\sim 0.44-0.48$. The phase
transition from deceleration to acceleration of the Universe occurs at the redshift
$z_{q=0}\sim 0.47-0.51$ in $68.3\%$ confidence level limits, as shown in figure
\ref{Fig5}. For GTF as unification of dark matter and dark energy, $-q_{z=0}\sim
0.50-0.61$ and $z_{q=0}\sim 0.49-0.68$ in $68.3\%$ confidence level limits as shown in
figure \ref{Fig9}. All these results are comparable with that estimated from 157 gold
data ($z_{\rm t}\simeq0.46\pm0.13$) \cite{Rie04}, but less than that obtained from
gold+SNLS SNIa data for DGP brane ($z_{q=0}\sim 0.8-0.93$) \cite{Guo06}.

\begin{figure}
\includegraphics[width=10cm]{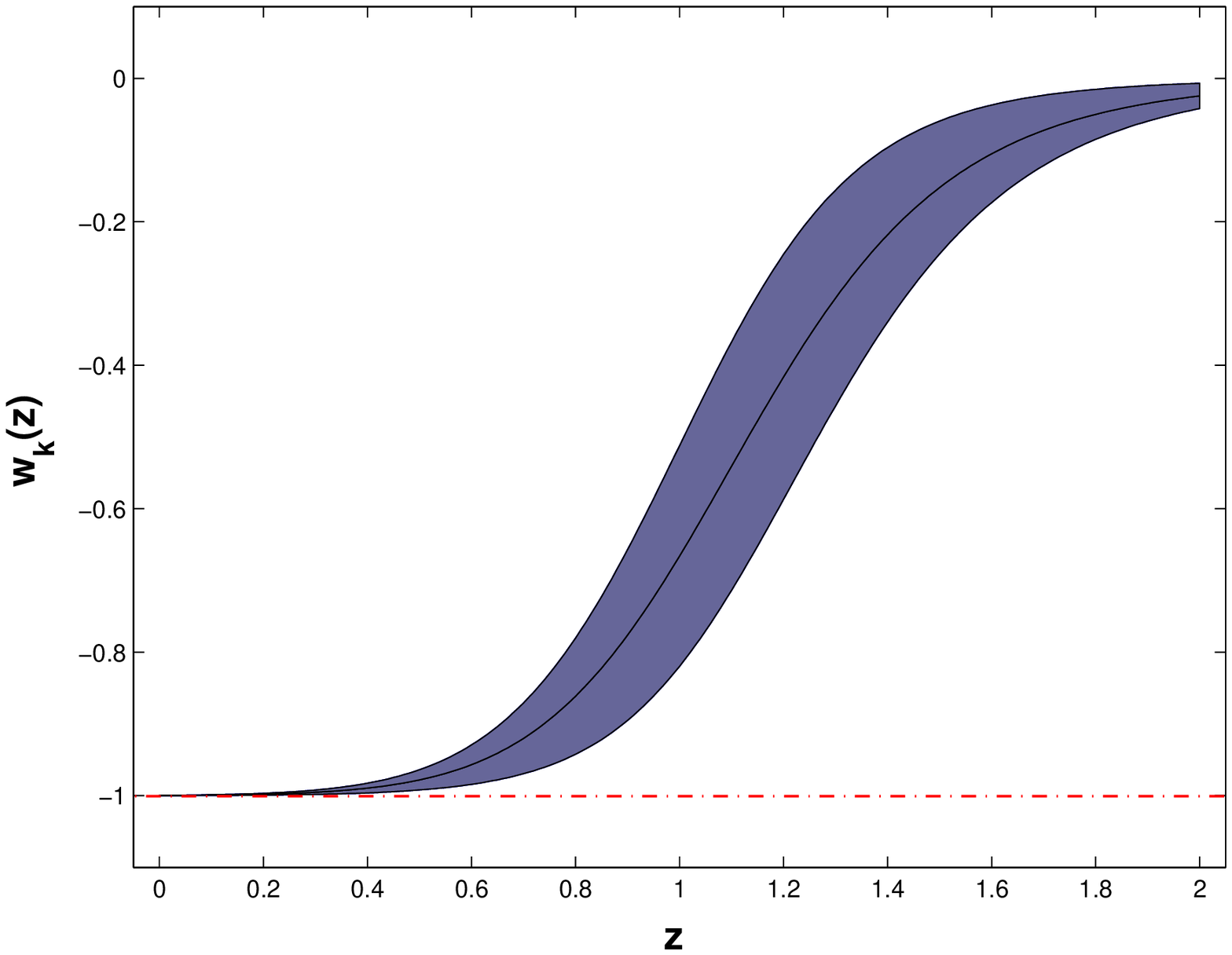}
\caption{The evolution of the equation of state parameter of GTF as
dark energy component only in $68.3\%$ confidence level limits from
fitting the combined data, compared with the $\Lambda$CDM case (the
dash-dot line). \label{Fig6}}
\end{figure}

\begin{figure}
\includegraphics[width=10cm]{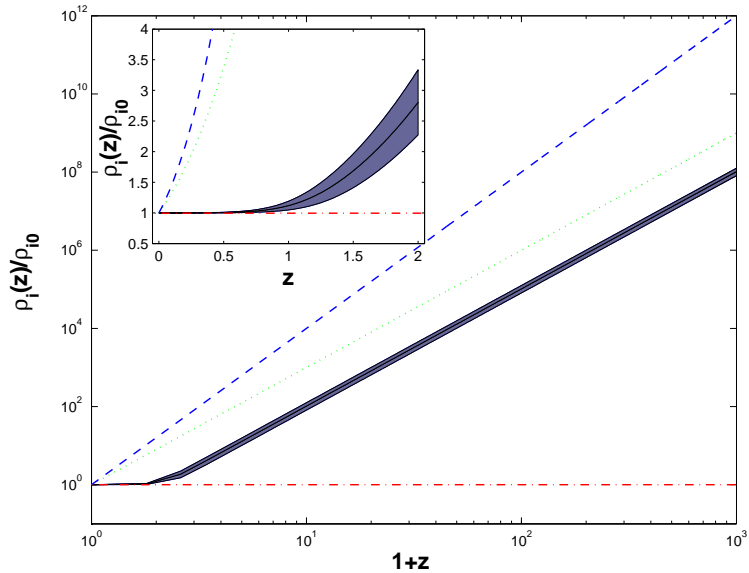}
\caption{The evolution of the energy density ratio of the GTF as
dark energy component only in $68.3\%$ confidence level limits from
fitting the combined data, compared with the cases of the radiation
(the dash line), the dark matter (the dot line), and the vacuum
energy (the dash-dot line). \label{Fig7}}
\end{figure}

\begin{figure}
\includegraphics[width=10cm]{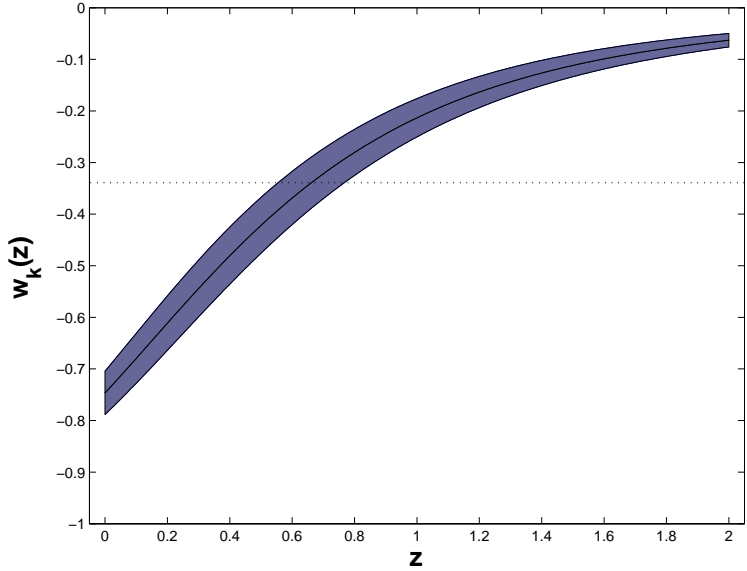}
\caption{The evolution of the equation of state parameter of GTF as
unification of dark matter and dark energy in $68.3\%$ confidence
level limits from fitting the combined data. \label{Fig10}}
\end{figure}

\begin{figure}
\includegraphics[width=10cm]{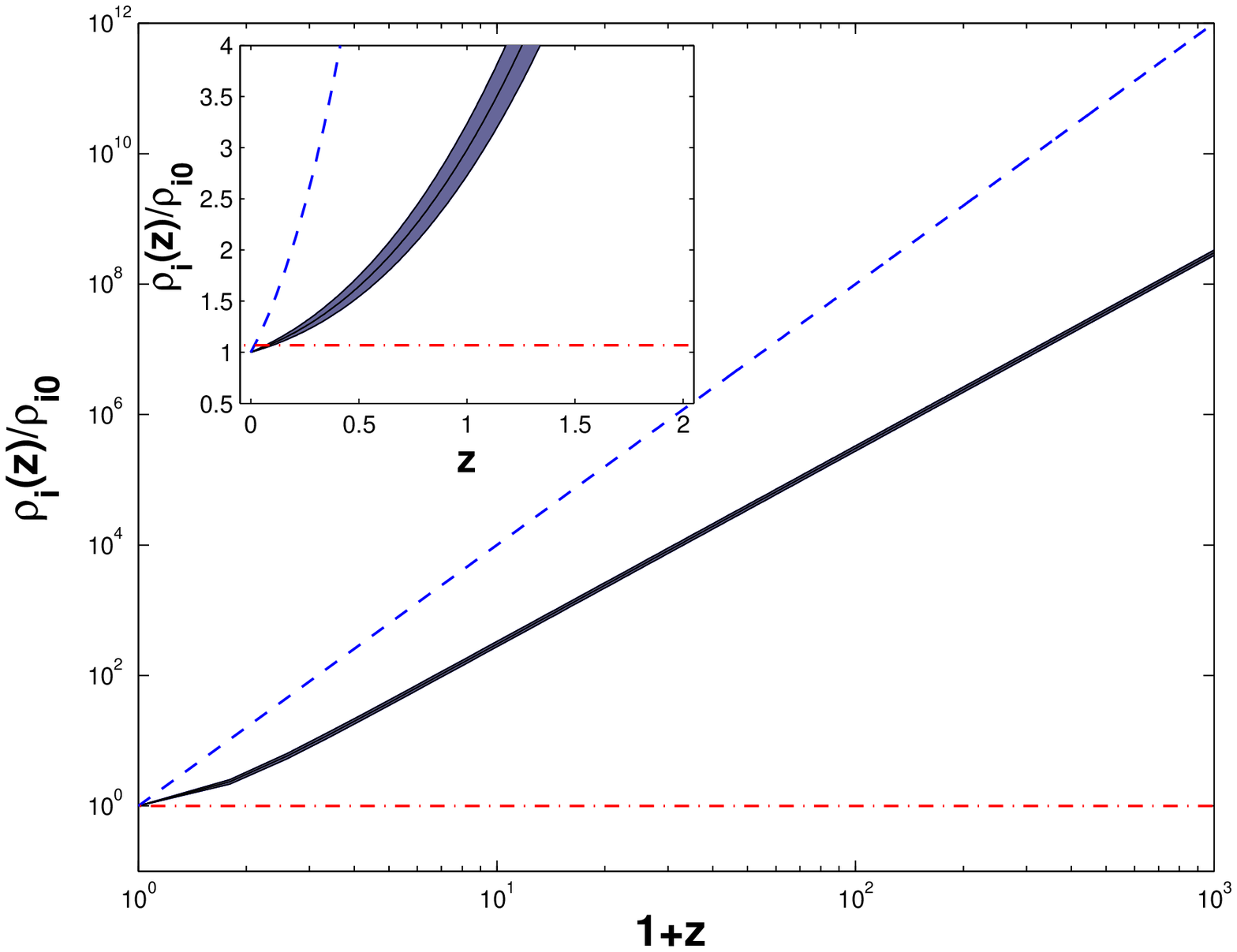}
\caption{The evolution of the energy density ratio of the GTF as
unification of dark matter and dark energy in $68.3\%$ confidence
level limits from fitting the combined data, compared with the cases
of the radiation (the dash line) and the vacuum energy (the dash-dot
line). \label{Fig11}}
\end{figure}

For the case of GTF as dark energy component only, figure \ref{Fig6}
and \ref{Fig7} show the evolution of the EoS parameter and the
energy density ratio of GTF dark energy at low or high redshift,
compared with the vacuum energy in both cases. For $z\gtrsim 2$, the
EoS parameter runs closely to $-0$, meaning the negative pressure of
the GTF dark energy approaches to zero rapidly, compared with the
cases of the radiation and the dark matter. Such behavior can, to a
certain degree, solve the fine-tuning problem \cite{Arm00, Chi}. For
GTF as unification of dark matter and dark energy, figure
\ref{Fig10} and \ref{Fig11} show the evolution of the EoS parameter
and the energy density ratio at low or high redshift, compared with
the cases of the radiation and the vacuum energy. All these results
at low redshift are consistent with that obtained in Ref.
\cite{wang} by model-independent methods in $68.3\%$ confidence
level limits.

\section{Growth of linear density perturbations}
Stability properties of some perfect fluid cosmological models are
studied extensively \cite{Bruni}, such as Refs. \cite{Gorini05,
Sandvik, Perrotta, Amendola05, Zimdahl, Avelino} concentrated on the
stability of GCG as unification of dark matter and dark energy,
Refs. \cite{Bean2003, Multamaki, Sen05} on the stability of GCG as
dark energy component only, and Refs. \cite{Frolov, Jain, Abramo} on
the stability of tachyon field dark energy.

\subsection{The case of GTF as dark energy only}
\begin{figure}
\includegraphics[width=10cm]{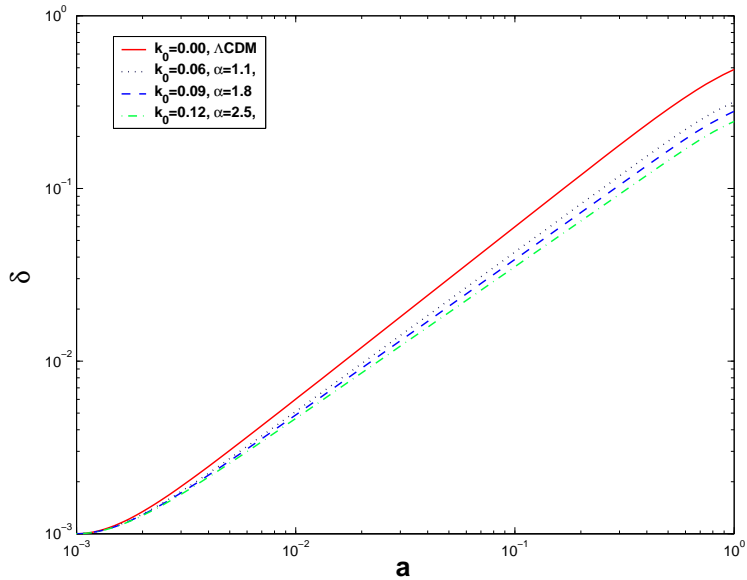}
\caption{The evolution of the matter density perturbation $\delta$
as a function of the scale factor $a$ (normalized to $a=1$ at the
present) for some selected values of the parameters ($k_0$ and
$\alpha$) of the GTF as dark energy in $68\%$ confidence level with
$\Omega_m=0.37$. \label{Fig12}}
\end{figure}

In this subsection, we study the growth of density perturbations for
the mixture of a matter fluid and a GTF dark energy fluid in the
linear regime on subhorizon scales. Assuming the GTF dark energy to
be a smooth, unclustered component (the only effect of the GTF
evolution is to alter the growth of matter perturbations through the
the effect of the GTF energy density on the expansion of the
universe), the growth equation for the linear matter density
perturbation, $\delta\equiv\delta\rho_{\rm m}/\rho_{\rm m}$, is
given by \cite{Multamaki, Sen05}
\begin{eqnarray}
\label{22}\delta''+\left(2+\frac{\dot{H}}{H^2}\right)\delta'+3c_1\delta=0,
\end{eqnarray}
where ``prime" denotes the derivative with respect to $\ln a$, ``dot" denotes the
derivative with respect to $t$, $H$ is the Hubble parameter for the background
expansion gives in Eq. (\ref{5}), and $c_1$ is given by
\begin{eqnarray}
\label{23}c_1=-\frac{1}{2}\frac{\Omega_{\rm m}}{\Omega_{\rm
m}+\Omega_{\rm k}[1+w_{\rm k0}(a^{6\alpha}-1)]^{1/2\alpha}},
\end{eqnarray}
with $w_{\rm k0}=1/(1+2k^{2\alpha}_0)$. For $k_0=0$, the equation
reduces to that for the $\Lambda$CDM model. The initial conditions
are chosen such that at $a=10^{-3}$, the standard solution
$\delta\sim a$ for Einstein-deSitter universe is reached. We have
integrated Eq. (\ref{22}) numerically from $a=10^{-3}$ to $a=1$ for
some selected values of the parameters ($k_0$ and $\alpha$) in
$68\%$ confidence level. Figure \ref{Fig12} shows the behavior of
$\delta$ as a function of the scale factor. Compared to the
$\Lambda$CDM universe, fluctuations grow more slowly in a universe
where GTF dark energy plays a role. For parameters ($k_0$ and
$\alpha$) changing in $68\%$ confidence level, $\delta$ deviates
slightly, consistent with the growth of linear density
perturbations. The behavior of $\delta$ in Fig. \ref{Fig12} agrees
with the result obtained in Ref. \cite{Sen05} in the framework of
GCG dark energy.

\subsection{The case of GTF as unification of dark matter and dark energy}
\begin{figure}
\includegraphics[width=10cm]{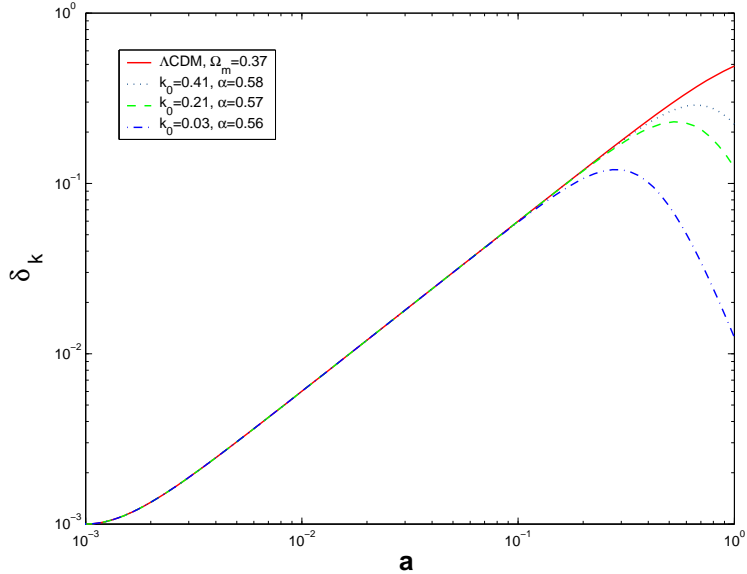}
\caption{The evolution of the GTF (unified dark matter) density
perturbation $\delta_{\rm k}$ as a function of the scale factor $a$
(normalized to $a=1$ at the present) for some selected values of the
parameters ($k_0$ and $\alpha$) in $68\%$ confidence level, compared
with the evolution of the matter density perturbation $\delta$ in
the case of $\Lambda$CDM. \label{Fig14}}
\end{figure}

Because baryons play a crucial role in the context of unified dark
matter/dark energy models \cite{beca03, Gorini07}, here we study the
growth of density perturbations for the mixture of a baryonic fluid
and a GTF fluid unifying dark matter and dark energy. In the
comoving synchronous gauge the relativistic equations governing the
evolution of perturbations in a two fluid (baryon and GTF) system
are \cite{beca03, Veer}
\begin{eqnarray}
\delta''_{\rm b}+\left(2+\frac{\dot{H}}{H^2}\right)\delta'_{\rm b}+
\frac{3}{2}\left[\Omega_{\rm b}\delta_{\rm b}+(1-3(2\alpha-1)w_{\rm k})\Omega_{\rm k}\delta_{\rm k}\right]=0,\\
\delta'_{\rm k}+(1+w_{\rm k})[\theta_{\rm k}/aH-\delta'_{\rm
b}]-6\alpha w_{\rm k}\delta_{\rm k}=0,\\
\theta'_{\rm k}+[1+3(2\alpha-1)w_{\rm k}]\theta_{\rm
k}+\frac{(2\alpha-1)w_{\rm k}k^2}{aH(1+w_{\rm k})}\delta_{\rm k}=0,
\end{eqnarray}
where $\delta''_{i}$ is the density contrast of the $i$th fluid
obeying $p_i=w_i\rho_i$, $\theta_{\rm k}$ is element velocity
divergence. Given $w_{\rm k}$ and $H$ as functions of $a$ we can
easily transform this set of equations into four first order
differential equations and integrate them using numerical method.
Since in the linear regime and deep into the matter era
$\delta_{i}\propto a$ implying $\delta'_{i}\propto a$ with
normalized initial conditions $[\delta_{\rm b}, \delta'_{\rm b},
\delta_{\rm k}, \theta]=[0.001, 0.001, 0.001, 0]$ for $a=0.001$ and
a prior $k=100h$ Mpc$^{-1}$ which corresponds to a scale of order
$50h^{-1}$kpc. Figure \ref{Fig14} shows the behavior of $\delta_{\rm
k}$ as a function of the scale factor. The fluctuations of GTF
density grow more slowly for $a\rightarrow1$, meaning GTF does not
behave as classical $\Lambda$CDM scenarios. The reason is that
baryons can carry over gravitational clustering when the GTF fluid
starts behaving differently from CDM \cite{beca03}.

\section{conclusions and discussions}
Assuming that the Universe is spatially flat, we place observational
constraints on GTF scenario with 182 gold SNIa data and two cosmic
microwave background parameters (the shift parameter and the
acoustic scale). For GTF as dark energy component only, the best-fit
values of the parameters at $68.3\%$ confidence level are:
$\Omega_{\rm m}=0.37\pm0.01$, $k_0=0.09^{+0.04}_{-0.03}$ and
$\alpha=1.8^{+7.4}_{-0.7}$ with $\chi^2_{\rm k,min}=159.30$
($p(\chi^2>\chi^2_{\rm k, min})=0.88$), comparing with $\chi^2_{\rm
\Lambda min}=168.59$ ($p(\chi^2>\chi^2_{\rm \Lambda, min})=0.77$) in
the $\Lambda$CDM case. For GTF as unification of dark matter and
dark energy, the best fit values of the parameters at $68\%$
confidence level are: $k_0=0.21^{+0.2}_{-0.18}$ and
$\alpha=0.57\pm0.01$, with $\chi^2_{\rm k, min}=167.27$
($p(\chi^2>\chi^2_{\rm k,min})=0.78$). In both cases, GTF evolves
like dark matter in the early universe.

To consider the best-fit models, we apply model-comparison
statistics. Comparing with GTF dark energy scenario, the combined
data do not support the $\Lambda$CDM case according to F-test and
AIC, but possibly support the $\Lambda$CDM case according to BIC.
Similarly the case of GTF as unification of dark matter and dark
energy is not supported according to F-test, AIC and BIC. Tested
with independent 9 $H(z)$ data points, GTF dark energy scenario is
favored over the $\Lambda$CDM model, and the $\Lambda$CDM model is
favored over GTF as unification of dark matter and dark energy. This
supports theoretical arguments against unifying dark matter and dark
energy into one scalar field. Of course, new and better data are
still needed to further discriminate between these models.

By investigating the deceleration parameter, we find that the
present value of the deceleration parameter $q(z)$ is $-q_{z=0}\sim
0.44-0.48$, the phase transition from deceleration to acceleration
of the Universe occurs at the redshift $z_{q=0}\sim 0.47-0.51$ in
$68.3\%$ confidence level limits for GTF as dark energy component
only; and $-q_{z=0}\sim 0.50-0.61$ and $z_{q=0}\sim 0.49-0.68$ in
$68.3\%$ confidence level limits for GTF as unification of dark
matter and dark energy. These results can be tested with future
cosmological observations. If assumed to be a smooth component, GTF
as dark energy component is consistent with the growth of linear
density perturbations. If GTF unifies dark matter and dark energy,
because baryons can carry over gravitational clustering when the GTF
fluid starts behaving differently from CDM, the growth of GTF
density fluctuations grow more slowly for $a\rightarrow1$, meaning
GTF do not behave as classical $\Lambda$CDM scenarios.

\begin{acknowledgments}
We thank Yun Wang, Zu-Hui Fan, Hao Wei, Pu-Xun Wu, Yan Wu, Wei-Ke
Xiao, Jian-Feng Zhou, Zhi-Xing Ling, and Bi-Zhu Jiang for
discussions. The anonymous referee is thanked for his/her patience
in reviewing this manuscript several times, as well as providing
insightful and constructive criticisms and suggestions, which
allowed us to improve the manuscript significantly. This study is
supported in part by the Ministry of Education of China, Directional
Research Project of the Chinese Academy of Sciences under project
No. KJCX2-YW-T03 and by the National Natural Science Foundation of
China under project no. 10521001, 10733010 and 10725313.
\end{acknowledgments}

\clearpage
\end{document}